\documentclass[pra,aps,floats,onecolumn,superscriptaddress,floatfix]{revtex4}

\usepackage{graphicx,amssymb,amsmath,ifthen}
\usepackage[active]{srcltx}
\usepackage[subrefformat=parens,labelformat=parens]{subfig}
\usepackage{lipsum}
\usepackage{physics}
\begin{document}
\title{Optical States in a 1-D Superlattice with Multiple Photonic Crystal Interfaces}  
\author{Nicholas J. Bianchi}
\affiliation{
 Department of Physics,
University of Rhode Island,
Kingston RI 02881, USA}
\author{Leonard M. Kahn}
\affiliation{
 Department of Physics,
University of Rhode Island,
 Kingston RI 02881, USA}

\begin{abstract}
Interface states in a 1-D photonic crystal heterostructure with multiple interfaces are examined. The heterostructure is a periodic network consisting of two different photonic crystals. In addition, the two crystals themselves are periodic, with one being made of alternating binary layers and the other being a quaternary crystal with a tunable layer. The second crystal can thus be smoothly transformed from one binary crystal to another. All individual photonic crystals in the superstructure have symmetric unit cells, as well as identical periods and optical path lengths. Therefore, as the tunable layer in the quaternary crystal expands, other layers will shrink.  It is found that the behavior of the localized modes in the band gaps is dependent on whether there is an even or odd number of interfaces in the heterostructure. With certain sequences of all dielectric photonic crystals, topological states are shown to split in two, whereas for other heterostructures they are shown to vanish. Additional resonant modes appear depending on how many crystals are in the heterostructure. If the tunable layer is frequency dependent, the band gap can still support topological/resonant modes with some band gaps even supporting two separate groups.
\end{abstract}
\maketitle

\section{Introduction}
 A photonic crystal (PC) is a periodic array of dielectrics and/or conductors used to scatter light \cite{Yablonovitch,John}. In a similar manner to how semiconductors control the passage of electrons, PCs possess passbands which allow photons in certain frequency ranges to propogate through the crystal and photonic band gaps (PBGs), which inhibit photon flow, producing regions of suppressed transmission. The existence of these pass and stop bands are governed by Bloch's Thereom. Photonic heterostructure devices are comprised of multiple periodic components that can produce transmission properties and field localization not seen in isolated crystals \cite{Istrate1, Istrate2}. Heterostructures with a single PC interface have been extensively studied.  Examples of localized behavior are the surface or interface modes, also known as optical Tamm \cite{Tamm} states (OTSs). These modes can exist at a boundary only if their field amplitudes decay away as the distance from the boundary increases in either direction. This means the wavevectors must be imaginary. In the case of a PC, this occurs if the mode is trying to travel through a PBG. These modes have been found in a variety of photonic structures including 1-D \cite{Kavokin,Vinogradov1,Vinogradov2,Gao} and 2-D \cite{Lin} PC interfaces, air-PC surfaces at oblique angles \cite{Feng}, and PCs bordering media with a graded refractive index \cite{Zheng}. Tamm states have also been investigated in various systems containing a PC with a tunable cap layer adjacent to a uniform medium. Examples include PCs containing superconducting layers \cite{Abouti}, systems containing metamaterials, both the PC layers \cite{Wang,Barvestani} and the uniform medium \cite{Namdar}, and  systems with liquid crystal \cite{Hajian} and chiral  \cite{Bashiri} cap layers. Note that in Ref. \cite{Feng}, despite the PC being adjacent to a uniform medium with positive dielectric constant, surface modes can still form due to total internal reflection. The component of the wavevector parallel to the boundary, $k_\parallel$, is large enough to cause the normal component, $k_\bot$ to become imaginary.
\begin{equation}
k_\bot = \sqrt{k^2 - k_\parallel^2} \label{wavevector}
\end{equation}
A varient of OTSs are the Tamm plasmon-polaritons (TPP) formed at a boundary between a metal and a PC \cite{Kaliteevski,Brand,Zhou}. In order for a TPP to form, the condition,
\begin{equation}
r_{\text{metal}}r_{\text{PC}}=1 \label{reflection}
\end{equation}
must be satisfied. The reflection coefficent $r_{\text{metal}}$ describes the amplitude of the electric field, incident from the PC side of the interface, reflecting off the metallic surface. In the same manner, $r_{\text{PC}}$ describes the electric field amplitude from a wave incident from the metallic side reflecting off the PC surface of the interface. In the case described in Ref.~\cite{Kaliteevski}, the TPP is excited at a frequency below the plasma frequency of the metal, implying that $r_{\text{metal}}=-1$. Therefore, to ensure that Eq.~\ref{reflection} remains satisifed, $r_{\text{PC}}=-1$, implying that the higher index material in the PC should be adjacent to the metal. In Ref.~\cite{Brand}, the plasmon is produced above the plasma frequency. Since the permittivity of the metal is now positive, $r_{\text{metal}}$ flips sign. For the state to exist now, the sign of $r_{\text{PC}}$ must also flip, meaning that, in the PC, the low index material is adjacent to the material. Similar to Ref.~\cite{Feng}, the state is supported on the metallic side by total internal reflection.

If an interface is generated between two PCs with symmetric unit cells, localized states at the boundary can form that are governed by the bulk band structure of the two crystals. These states are referred to as topological interface states. Xiao \textit{et.al.} \cite{Xiao} showed that their existence in a PBG can be predicted by ensuring that the imaginary parts of the surface impedances for the two crystals sum to zero in the selected gap. Their work established a relation between the sign of the impedance $Z$ for a PBG and sum of all Zak \cite{Zak} phases, $\theta^{\text{zak}}_m$, below the gap, where $m$ denote the (isolated) bands,
\begin{equation}
\text{sign}(\text{Im}(Z^{(n)})) = (-1)^{n+l}\exp \left(i \sum_{m=1}^{n} \theta^{\text{zak}}_m \right) \label{Zak}
\end{equation}
In Eq.~\eqref{Zak}, $n$ is the PBG where the impedance is calculated and $l$ denotes the number of points where two bands cross below band gap $n$. Due to the PC unit cells possessing inversion symmetry, all Zak phases can only take on the values of $\pi$ or $0$ \cite{Zak}, and thus provide a useful measure for identifying topological states. Band gap $n$ contains a topological state if $Z_L+Z_R=0$, where the subscripts indicate the PCs to the left/right of the interface. Through control of $\theta^{\text{zak}}_m$, topological states have been demonstrated in both 1-D \cite{Choi,Cai} and 2-D \cite{Yang} systems.

Heterostructures with multiple PC/PC or PC/metallic interfaces have more degrees of freedom due to the increased number of tunable parameters, as compared to a single interface system, leading to a much richer display of resonant states. Through the control of parameters within the heterostructure, several examples of coupling between resonant states have been demenstrated \cite{Zhou,Fei, Iorsh,Durach,Cox,Hu}. As an extension to the work in Ref.  \cite{Bianchi}, the behavior of interface states is investigated in a heterostructure consisting of alternating binary and quaternary PCs. If the number of binary and quaternary crystals in the stucture are the same, then there is an odd number of interfaces. In this case if the first PC in the heterostructure is binary (quaternary), then, after the alternating pattern, the last will be quaternary (binary). The orginal topological state from the two crystal hetrostructure remains but is now accompanied by a sequence of resonant states on either side. The total number of states, including the original, is equal to the number of interfaces. For an even number of interfaces, there are two possible configurations. One possibility is to have the first and last PC of the structure be binary. In this case, it is found that the orginal topological state vanishes while the resonant states remain. The other case is to have the first and last crystals be quaternary. With only a single binary PC sandwiched between two quaternary PCs, the topological state splits. If more layers are added in this scenario, keeping the two ends quaternary, the split state is joined by resonant states. 

\begin{figure}[!ht]
	\centering
	\subfloat{\includegraphics[width=0.6\columnwidth]{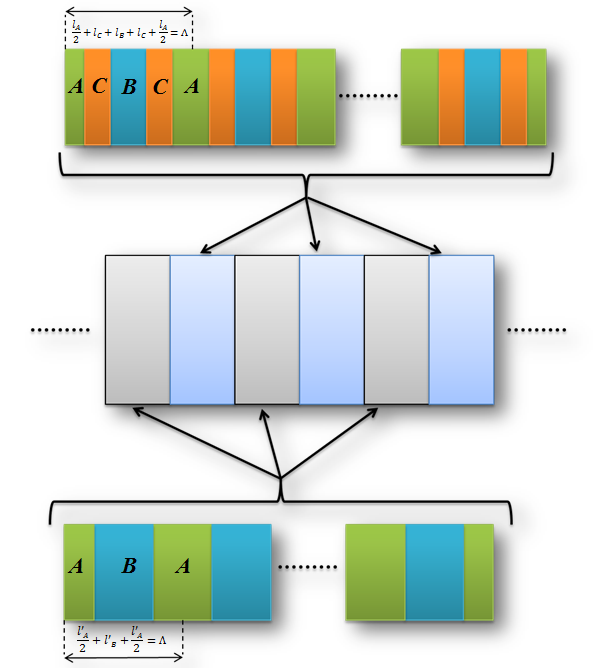}}
	\caption{Schematic of PCs and overall heterostructure. The heterostructure is displayed in the middle as repeating light grey and blue "slabs". Each of these "slabs" represent entire photonic crystals. The grey slabs represent binary PCs, displayed at the bottom. In the bottom diagram, dark green is layer $A$ and dark blue is layer $B$. The vertical dashed lines and double arrow show one unit cell. The primes indicate that the individual layer lengths are different from those of the quaternary PC. Note that the PC is capped on both sides with a half-width of layer $A$, making the unit cells symmetric. The top diagram shows the quaternary PC, which is represented in the middle picture as the light blue regions. In the quaternary PC, layers $A$ and $B$ are the same material as in the binary crystal. Layer $C$ is orange. As in the binary case, the  symmetric unit cell is displayed by the double arrow.}
	\label{PCM}
\end{figure}

\section{Methods}

Our work was conducted using transfer matrix method (TMM) \cite{Yariv}. Keeping with Ref.~\cite{Bianchi}, all variables are made dimensionless for convenience. The lengths of the individual PC layers, $l_i$, are scaled to the unit cell period, $\Lambda$: $d_i=l_i/\Lambda$ and are such that $\Lambda$ and the optical path, $\Gamma$, for a unit cell are constant. In the heterostructure, shown as the middle image in Fig.~\ref{PCM}, the periods for all the individual PCs are equal, as are the optical paths. The binary PCs are the gray regions and the quaternary PCs are the light blue regions. Since there is no fixed length scale, we set $\gamma=\Gamma/\Lambda$. For the quaternary PC, shown at the top of Fig.~\ref{PCM}, the widths of layers $A$, in green, and $B$, in blue,  can be expressed in terms of a free parameter, the width of the introduced layer, $d_C$, in orange, \cite{Bianchi},

\begin{equation}
d_A = \frac{\gamma - n_B - 2(n_C - n_B)d_C}{n_A - n_B}   \label{eq:dA}
\end{equation}

\begin{equation}
d_B = \frac{\gamma - n_A - 2(n_C - n_A)d_C}{n_B - n_A}    \label{eq:dB}
\end{equation}
Note that $d_C$ can only take on values in which both Eqs.~\eqref{eq:dA} and \eqref{eq:dB} are non-negative. When $d_C$ reaches its maximum, the quaternary PC will become binary again, but with configuration, $...CBCBC...$, if $d_A$ tends to zero, or, $...ACACA...$, if $d_B$ tends to zero. For the special case, $\gamma=n_C$, both $d_A$ and $d_B$ will be zero when $d_C$ reaches its maximum; this will result in a uniform layer $C$.
The lengths of the layers in the binary PC, displayed at the bottom of Fig.~\ref{PCM}, are simply Eqs.~\eqref{eq:dA} and \eqref{eq:dB} but with $d_C=0$ and thus do not change. The index of refraction of layer $j$ is $n^2_j = \epsilon_j \mu_j$, where $\epsilon_j$ and $\mu_j$ are the (relative) permitivites and permeabilites. In the binary and quaternary crystals, the $n_A$'s are the same and the $n_B$'s are the same, although $n_A \neq n_B$ \cite{Bianchi}. 

For the system described in Fig.~\ref{PCM}, we only consider an electric field incident from the left, $E_{1+}$. The reflected field is $E_{1-}$ and the field that is transmitted through the entire structure is $E^{\prime}_{(N+1)+}$. To compute the transmission spectra for the system, first  we must construct the transfer matrix, $M$, from the individual interface matrices, $I_j$, and propagation matrices, $P_j$, where the index, $j$, specifies the layer in question \cite{Orfanidis},
\begin{equation}
I_j=
\frac{1}{\tau_j}
\begin{pmatrix}
1            &  r_j \\[6pt]
r_j    &   1        \\
\end{pmatrix}
\end{equation}

\begin{equation}
P_j=
\begin{pmatrix}
e^{2 \pi i n_j d_j \xi}    & 0                                         \\[6pt]
0                                    &   e^{-2 \pi i n_j d_j \xi}        \\
\end{pmatrix}
\end{equation}
where $r_j$ and $\tau_j$ are the reflection and transmission coeffiecents, respectively,
\begin{equation}
r_j = \frac{\mu_{j+1} n_{j} - \mu_{j} n_{j+1} }{\mu_{j+1} n_{j} + \mu_{j} n_{j+1} }
\end{equation}
\begin{equation}
\tau_j = \frac{2 \mu_{j+1} n_{j}  }{\mu_{j+1} n_{j} + \mu_{j} n_{j+1} }
\end{equation}
In scaled variables, the phase argument, $i k_j l_j$ becomes $2 \pi i n_j d_j \xi$. The frequency, $f$, becomes $\xi=f\Lambda/c_0$, where $c_0$ is the speed of light in vacuum. The incident and scattered field are related by,

\begin{equation}
\begin{pmatrix}
E_{1+}\\[6pt]
E_{1-}\\
\end{pmatrix}
=
\begin{pmatrix}
M_{11}  & M_{12} \\[6pt]
M_{21}  &   M_{22}           \\
\end{pmatrix}
\begin{pmatrix}
E^{\prime}_{(N+1)+}\\[6pt]
0\\
\end{pmatrix}
\end{equation}
where,
\begin{equation}
M = 
\begin{pmatrix}
M_{11}  & M_{12} \\[6pt]
M_{21}  &   M_{22}           \\
\end{pmatrix}
=
\prod_{j=1}^{N}I_j P_j I_{N+1}
\end{equation}
The transmitted power is calculated via,
\begin{equation}
T(\xi,d_C) = \abs{\frac{1}{M_{22}}}^2
\end{equation}


\section{Results}

In our first investigation, all layers of the heterostructures are assumed to be lossless dielectrics with no material dispersion. For both the binary and quaternary PCs, $\epsilon_A=6, \epsilon_B=\mu_A=\mu_B=1$. In the quaternary PC, $\epsilon_C=3$ and $\mu_C=1$. For simplicity, all PCs are given the same number of unit cells,$N_\Lambda$, periods and optical paths. In the following systems, $N_\Lambda=4$ and $\gamma=1.5$. With $d_C$ as a free parameter, $d_A$ and $d_B$ of the quaternary PC are described by Eqs.~\ref{eq:dA} \& ~\ref{eq:dB}. Since the physics of interface states is valid in any PBG, we will restrict ourselves to the $3^{\text{rd}}$ one since this is the lowest gap that produces such states with the described parameters. For convience, when describing individual PCs of the heterostructures, we will use $b$ for binary PC and $q$ for quaternary PC.

\begin{figure}[!ht]
	\centering
	\subfloat[][]{\includegraphics[width=0.33\columnwidth]{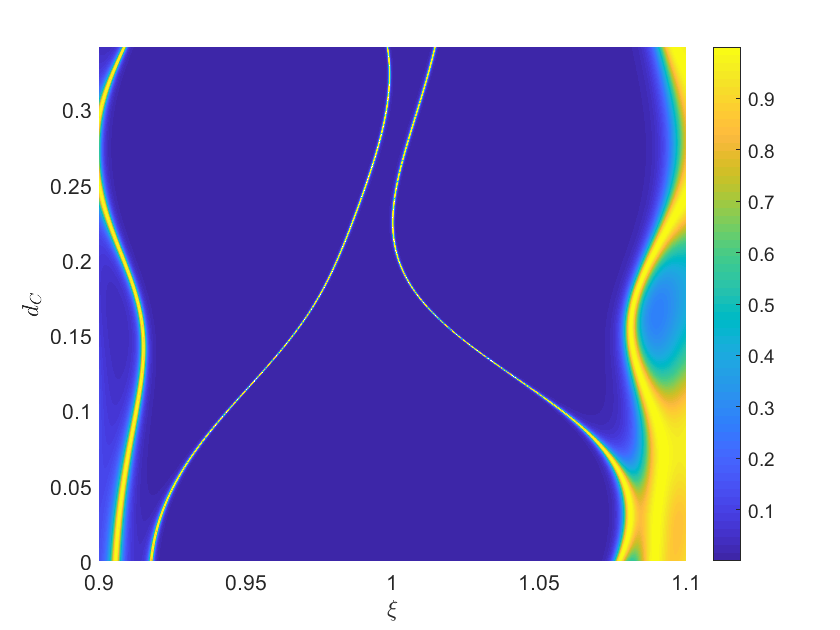}               \label{T_bqb}      }
	\subfloat[][]{\includegraphics[width=0.33\columnwidth]{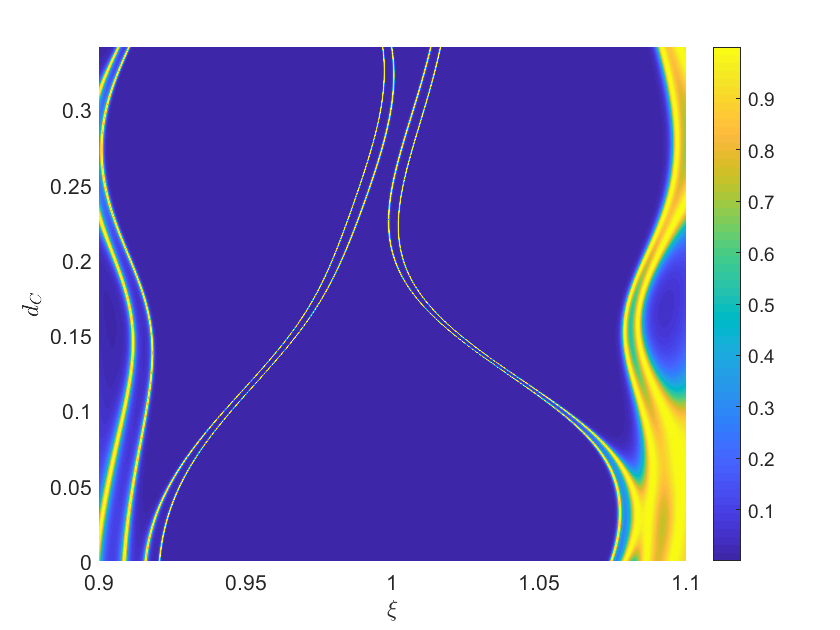}            \label{T_bqbqb}       }
	\subfloat[][]{\includegraphics[width=0.33\columnwidth]{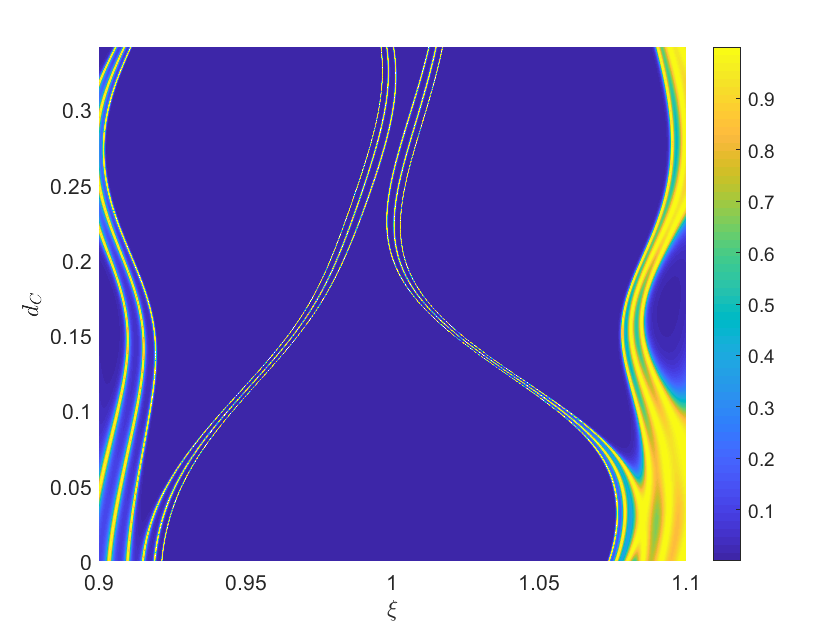}           \label{T_bqbqbqb}    }\\
	\subfloat[][]{\includegraphics[width=0.33\columnwidth]{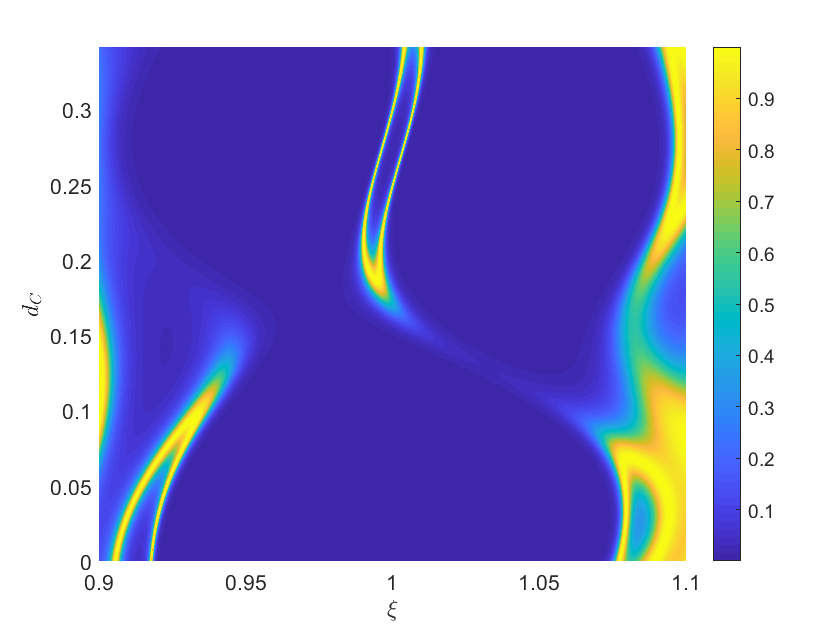}               \label{T_qbq}      }
	\subfloat[][]{\includegraphics[width=0.33\columnwidth]{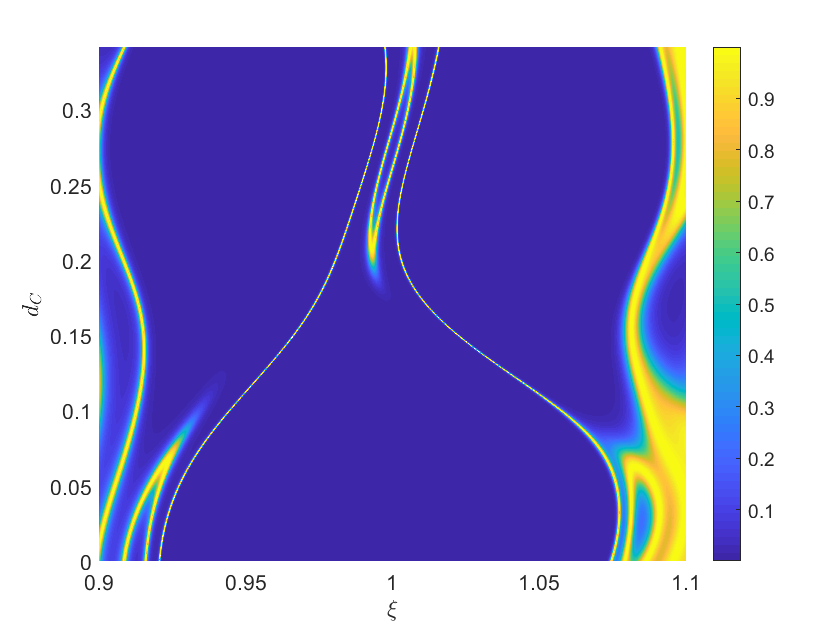}            \label{T_qbqbq}       }
	\subfloat[][]{\includegraphics[width=0.33\columnwidth]{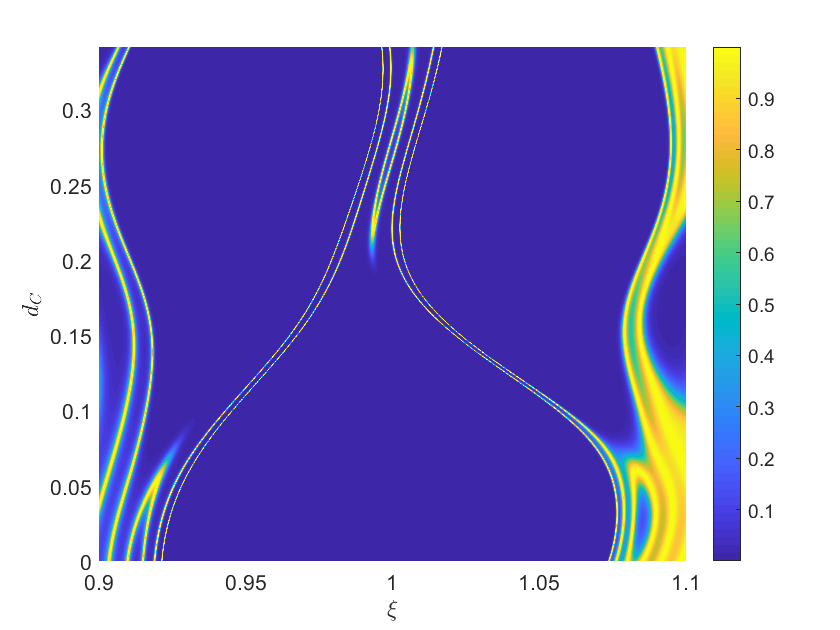}           \label{T_qbqbqbq}    }\\
	\subfloat[][]{\includegraphics[width=0.33\columnwidth]{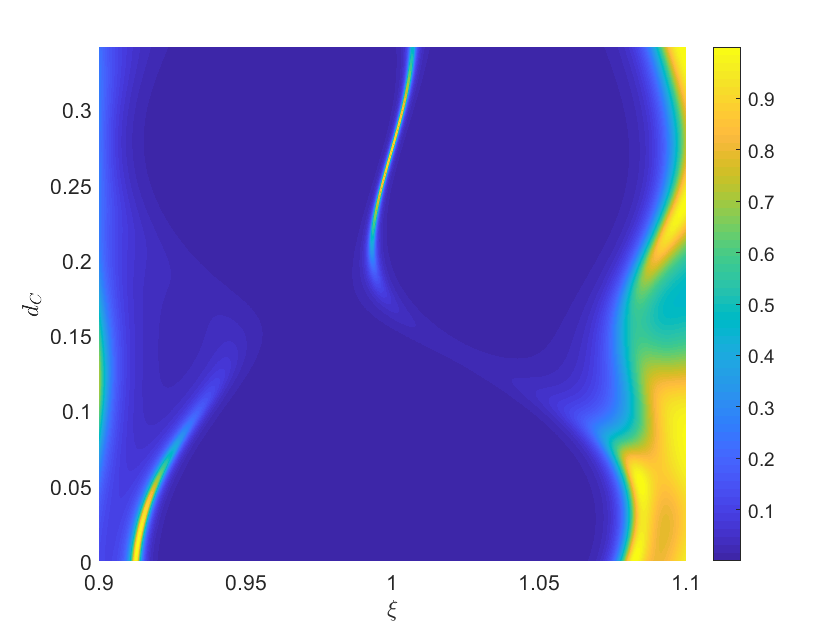}            \label{T_bq}       }
	\subfloat[][]{\includegraphics[width=0.33\columnwidth]{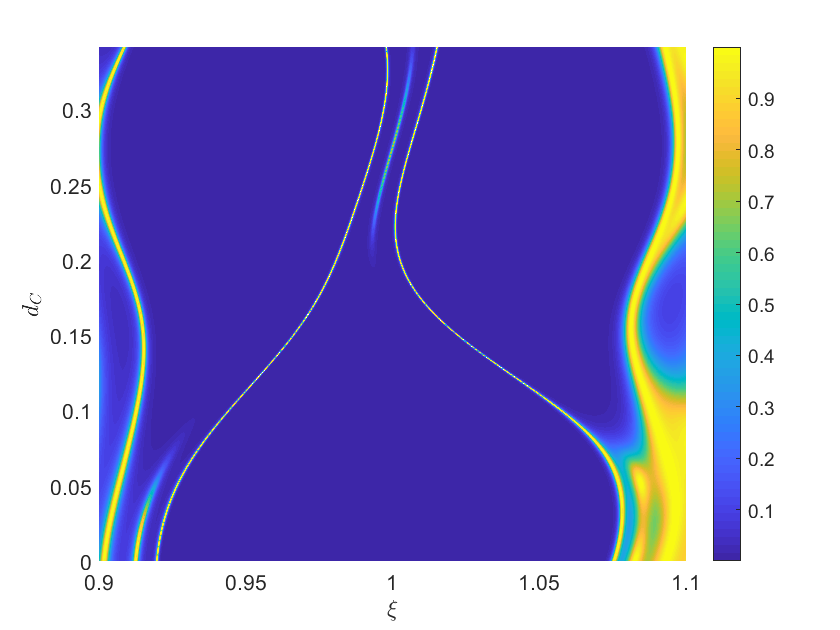}           \label{T_bqbq}    }
	\subfloat[][]{\includegraphics[width=0.33\columnwidth]{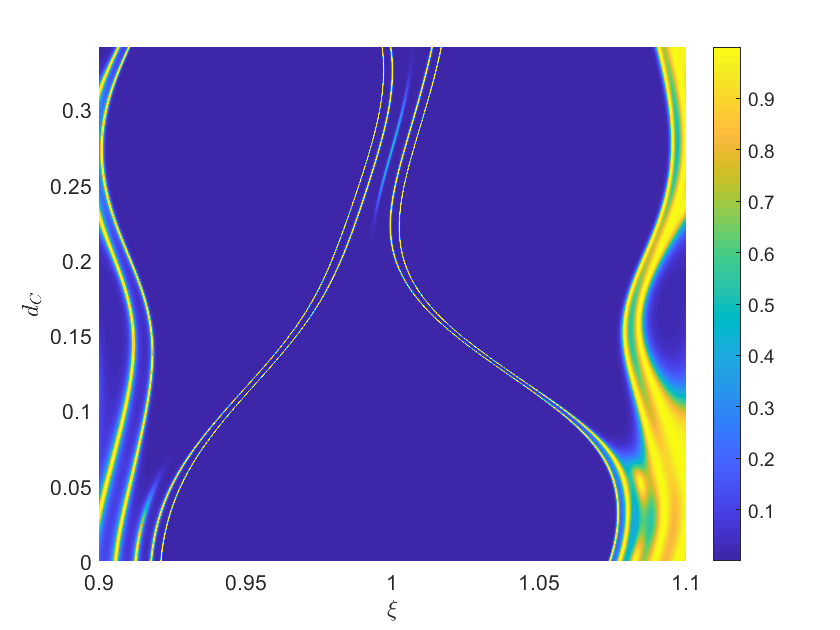}           \label{T_bqbqbq}    }
	\caption{Transmission map for PC heterostructure in $3^\text{rd}$ PBG. The stucture consists of alternating binary $(b)$ and quaternary $(q)$ PCs. Each PC has 4 symmetric unit cells. $\epsilon_A=6$, $\epsilon_C=3$, $\epsilon_B=\mu_A=\mu_B=\mu_C=1$ (a) $bqb$ (b) $bqbqb$ (c) $bqbqbqb$ (d) $qbq$ (e) $qbqbq$ (f) $qbqbqbq$ (g) $bq$ \cite{Bianchi} (h) $bqbq$ (i) $bqbqbq$}
	\label{Trans}
\end{figure}

Transmission through the heterosturcture depends on the specific configuration of the binary and quaternary PCs. Fig.~\ref{Trans} displays nine different transmission examples. In Figs.~\protect\subref*{T_bqb}-\protect\subref*{T_bqbqbqb}, the stucture is sandwiched between two binary PCs, while in Figs.~\protect\subref*{T_qbq}-\protect\subref*{T_qbqbqbq}, the two endlayers are quaternary PCs. In both cases, the number of interfaces from left to right is 2, 4, and 6. Since there is an even number of interfaces in the heterostructure, a single topologial peak (Fig.~\protect\subref*{T_bq}) is absent, even though Eq.~\ref{Zak} states that there is a change in the sign of surface impedance  between the binary and quaternary components as $d_C$ increases from 0 to 0.341. For the transmission maps in the top row, the heterostructure has the form $bqb$, $bqbqb$ and $bqbqbqb$. It can be seen that the state in a single interface system splits into two sets of resonances, with one set below the original frequency and the other above. At $d_C=0$, all these states exist as pass band modes; however, as $d_C$ increases, they begin to wander into the band gap. As these resonant states appear at all values of $d_C$, they are not topological in nature, although after the impedance for the quaternary layers flips sign (see Eq.~\ref{Zak}), they appear to cluster together and move in a   similar manner to the topological state in the single binary-quaternary interface heterostructure. The transmission for all these states remains at unity for all values of $d_C$.

\begin{figure}[!ht]
	\centering
	\subfloat[][]{\includegraphics[width=0.4\columnwidth]{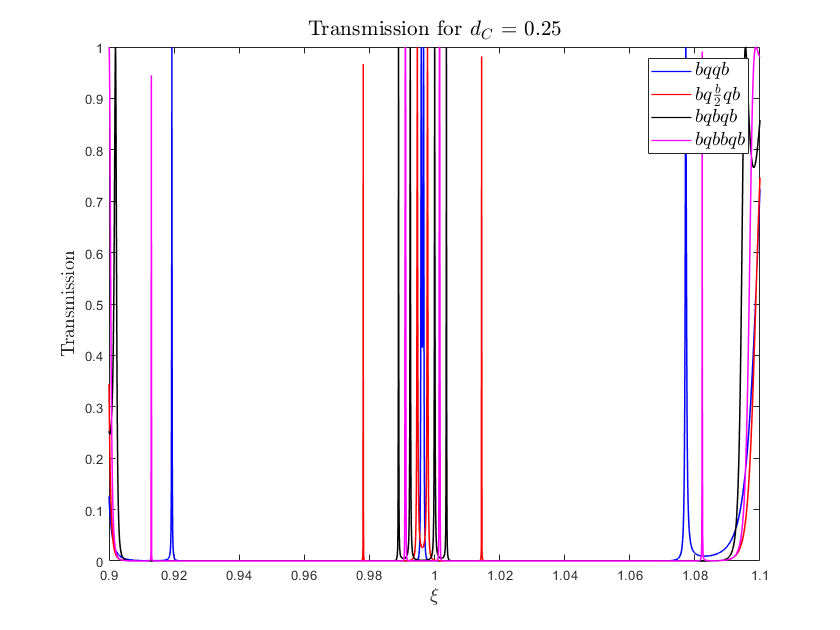}   \label{T1_dC} }
	\subfloat[][]{\includegraphics[width=0.4\columnwidth ]{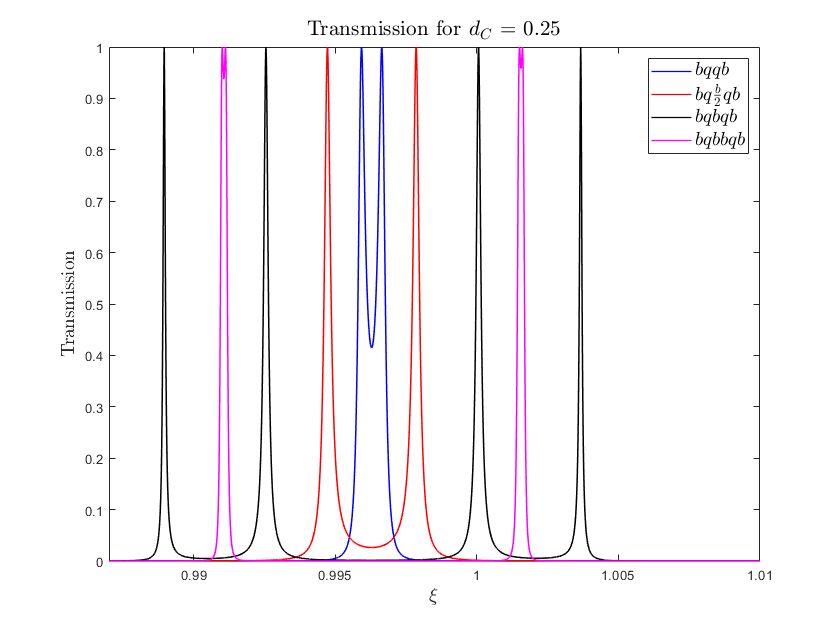}	      \label{T2_dC}}
	\caption{(a) Transmission resonance behavior for PC heterostucture $bq...qb$ as the number of central binary crystals varies. Each $b$ and $q$ 		                          represents 4 unit cells ($b/2$ is 2 unit cells). (b) Central region of (a).}
	\label{T_dC}
\end{figure}

When the heterostructure changes from $bqb$ to $bqbqb$, the two states themselves split into pairs such that these pairs (Fig.~\protect\subref*{T_bqbqb}) each have a higher and lower frequency state relative to their respective states in Fig.~\protect\subref*{T_bqb}. This splitting is illustrated in  Fig.~\ref{T_dC}. A horizontal slice of Fig.~\protect\subref*{T_bqbqb} at $d_C=0.25$ is considered, except now the states are plotted for varying thickness of the middle cystral. Each binary and quaternary represent 4 unit cells; $b/2$ represents 2 unit cells. In Fig.~\protect\subref*{T1_dC}, we see two distant edge states (blue) in the absence of a middle $b$: $bqqb$. To see the two interface states, though, we must zoom into the cluttered middle region. These interface states are clearly seen in  Fig.~\protect\subref*{T2_dC}. When two binary unit cells  are inserted in the center of the structure ($bq(b/2)qb$), there is now strong coupling between the two central states and the two edge states. The edge states rapidly move toward the central region. Inserting another two binary unit cells  produces the familiar structure $bqbqb$ and the black transmission profile. Doubling the central region causes coupling of the states in each pair to weaken due to the increased distance bewteen the interface pairs $bqb$. This is seen in the magenta curve as the four peaks mostly merge into two, recovering Fig.~\protect\subref*{T_bqb} at $d_C=0.25$. There is also a new pair of edge states in  Fig.~\protect\subref*{T1_dC}.

An important change occurs in the transmission behavior as $d_C$ increases if the sandwiching layers are quaternary. In Fig.~\protect\subref*{T_qbq}, the heterostructure is $qbq$ and a topological state is observed in the upper half of the map; however, it splits into two seperate peaks since there are two interfaces. With additional layers, the structure becomes $qbqbq$ and $qbqbqbq$, shown in Fig.~\protect\subref*{T_qbqbq} and  Fig.~\protect\subref*{T_qbqbqbq} respectively. Here the split topological state is strattled by resonant states that behave similarily to those described in  Fig.~\ref{T_dC}. To help understand why the topological state appears in the $qb...bq$ but not the $bq...qb$ configuration, it is helpful to examine the behavior of the band at small $d_C$. Recall that when Eq.~\ref{Zak} produces opposite signs for the isolated quaternary and binary PCs, the number of resonant states in the middle of the PBG must be equal to the number of interfaces in the composite heterostructure. For $bq...qb$, all the states remain separate (\textit{i.e.} states do not merge). For example, let's consider Fig.~\protect\subref*{T_bqbqbqb}. Since there are 6 interfaces, the 6 states that enter the  band gap are the 3 closest pass band states on either side of the gap. Compare this to Fig.~\protect\subref*{T_qbqbqbq}, where only the two closest states on either side of the PBG wander into the gap when $d_C$ increases from 0. The third closest states to the PBG are seen merging and disappearing with other states in adjacent bands on the far left and far right of the map. As $d_C$ continues to increase, there are temporarily only 4 states. Therefore in order to have a total of 6 states, the split topological state must appear after the phase transition. 

When the two endlayers are different, we get the sequence $bq...bq$. Figs.~\protect\subref*{T_bq}-\protect\subref*{T_bqbqbq} are $bq$, $bqbq$, and $bqbqbq$ respectively. Since there are now an equal number of binary and quaternary crystals in the structure, there are an odd number of interfaces and reversing the order of the components ($bq\rightarrow qb$) will not change the transmission. Fig.~\protect\subref*{T_bq} is the familiar single topological state from heterostruture $bq$. For Fig.~\protect\subref*{T_bqbq} \& \protect\subref*{T_bqbqbq}, the addition of $bq$ layers produces resonant states that behave like those discussed previously. 

\begin{figure}[!ht]
	\centering
	\subfloat{\includegraphics[width=0.4\columnwidth]{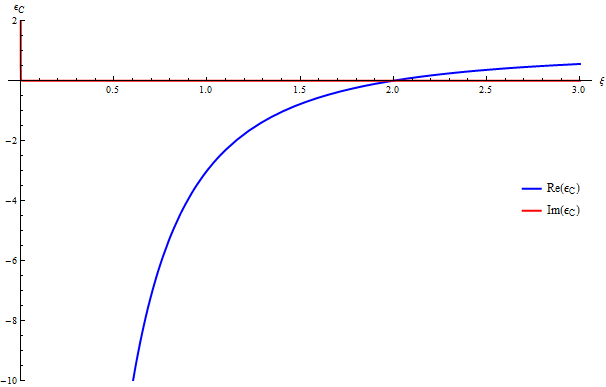}   }
	\caption{Real and imaginary parts of $\epsilon_C$ for metallic layer $C$. Plasma frequency $\xi_p=2$ and collision frequency $g=10^{-10}$. Frequencies are scaled according to $\xi = f\Lambda/c_0$.}
	\label{Epsilon}
\end{figure}

\begin{figure}[!ht]
	\centering
	\subfloat[][]{\includegraphics[width=0.4\columnwidth]{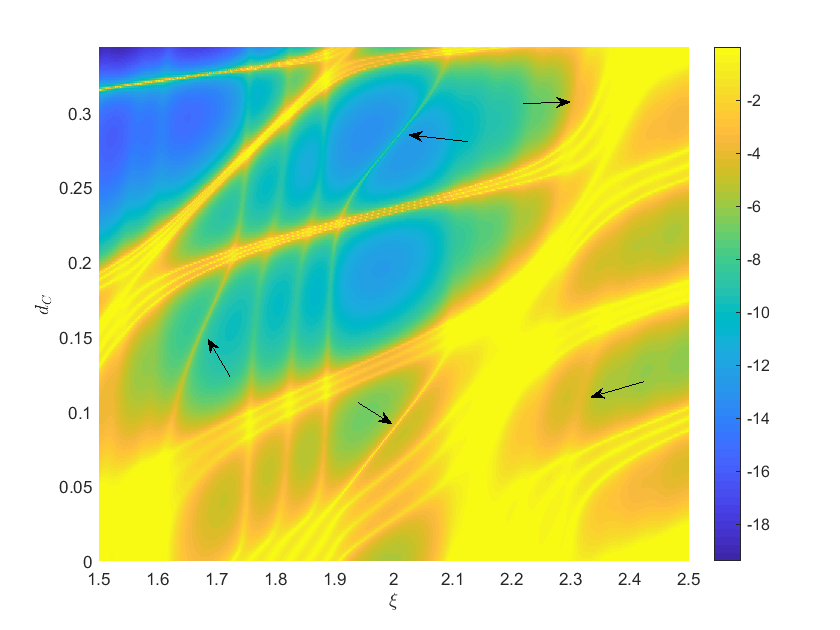}   \label{Trans_bq_M} }
	\subfloat[][]{\includegraphics[width=0.4\columnwidth ]{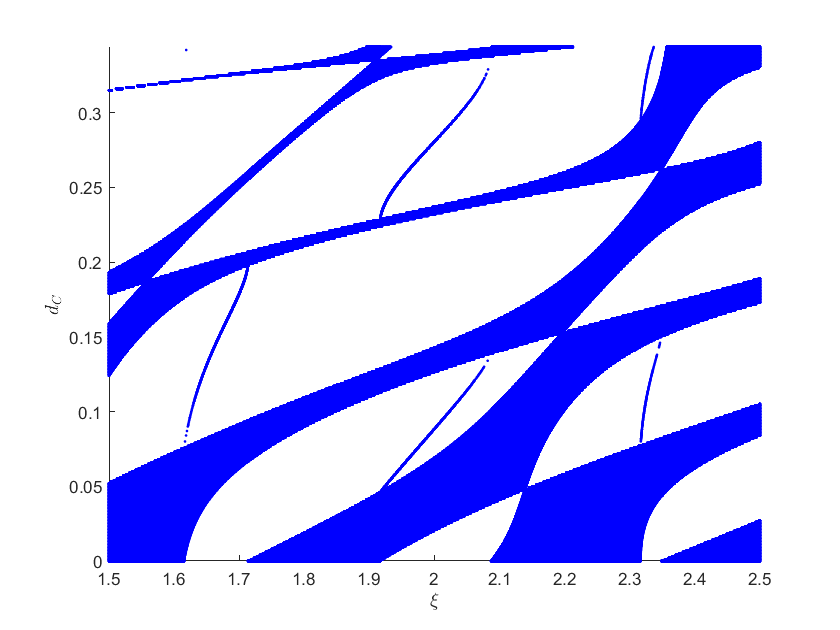}	      \label{Impedance_M}}
	\caption{(a) Transmission map behavior for PC heterostucture $bq$ where the quaternary PC contain metallic layer $C$.  $\epsilon_A=6$, $\epsilon_B=\mu_A=\mu_B=\mu_C=1$. $\epsilon_C$ is given by Eq.~\ref{Drude}.  Note the five topological states indicated by the arrows. The color scheme is logarithmic. (b) These states correspond to where the imaginary part of the impedance is zero.}
	\label{Trans_Impedance_bq_M}
\end{figure}

In our second investigation, layer $C$ is given a permittivity with frequency dependence, in accordance with the Drude model of dispersion,
\begin{equation}
\epsilon_C=1-\frac{\xi_p^2}{\xi^2+i g \xi}  \label{Drude}
\end{equation}
 where $\xi_p$ and $g$ are the dimensionless plasma and collision frequencies. Eq.~\ref{Drude} is plotted in Fig.~\ref{Epsilon} with plasma frequency $\xi_p=2$ and negligible collision frequency $g=10^{-10}$. Therefore, layer $C$ acts as a metal. Layers $A$ and $B$ remain unchanged. Since the optical path in metal is not constant with frequency, the layer width defined in Eqs.~\ref{eq:dA} \& \ref{eq:dB} are given simplier forms,
\begin{equation}
d_A = \frac{\gamma - n_B }{n_A - n_B}-d_C   \label{eq:dA_2}
\end{equation}

\begin{equation}
d_B = \frac{\gamma - n_A }{n_B - n_A}-d_C   \label{eq:dB_2}
\end{equation}
Now, $\gamma$ is only relevant when defining the layer widths before the metal is introduced. As $d_C$ increases, the band gap closing points are skewed towards higher frequencies due to the behavior of Eqs.~\ref{eq:dA_2} \& \ref{eq:dB_2}. As a concequence of this, toppological states in a single interface $bq$ system do not start and terminate at the closing points nor are they positioned near the center of the gap. An example of this behavior is shown in Fig.~\ref{Trans_Impedance_bq_M}. In  Fig.~\protect\subref*{Trans_bq_M}, the transmission map is plotted around $\xi=2$. The metallic layer, $d_C$
, follows the behavior in Fig.~\ref{Epsilon}. Note that we can have a case where one gap (top center) can support two states. The left state is much sharper than the right one. Also worth noting is that the two center states appear to cross the plasma frequency of the metallic layer without anything unusual happening. This is acceptable because the effective plasma frequency of the entire heterostructure is much lower than the plasma frequency of the metallic inclusion, so the effective permittivity of the heterosructure is positive in the region of these states \cite{Manzanares-Martinez}. This means that all visible gaps in Fig.~\protect\subref*{Trans_bq_M} are classifed as PBGs. There are also  two distinct groups of  Fabrey-Perot resonances. The brighter, more slanted triplets that largely encase the PBGs are caused by coupling among the 3 interfaces of the four unit cells in the quaternary PC.  There is also a fainter vertical triplet of resonances between about $1.72<\xi<1.9$, that is caused by the three interfaces in the binary PC. As $d_C$ increases, the leftmost topological state eventually appears to turn into one of these resonances and the rightmost of these states breaks away to become the top-center topological state. The equation $\text{Im}(Z_b+Z_q)=0$ is plotted in Fig.~\protect\subref*{Impedance_M}, showing the exact location of those five topological states.

\begin{figure}[!ht]
	\centering
	\subfloat[][]{\includegraphics[width=0.4\columnwidth]{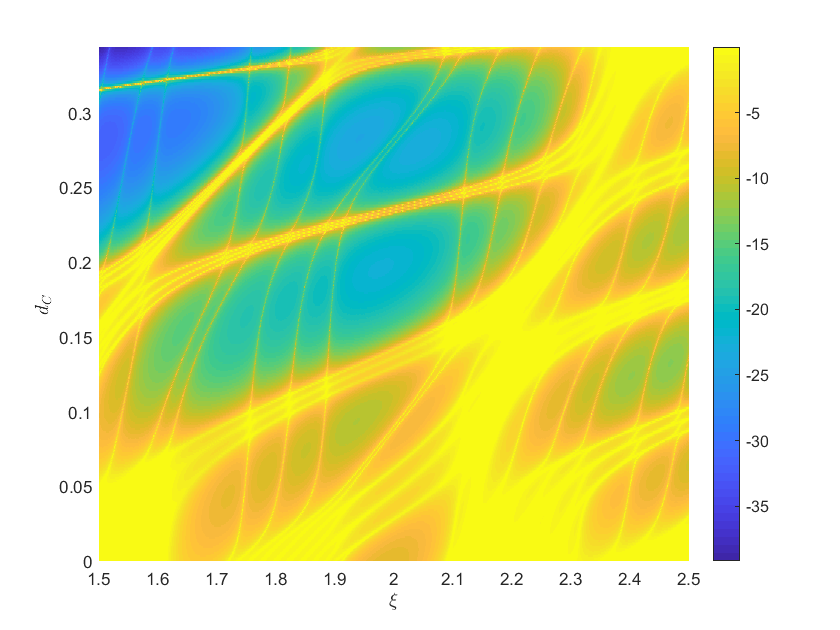}   \label{Trans_qbq_M} }
	\subfloat[][]{\includegraphics[width=0.4\columnwidth ]{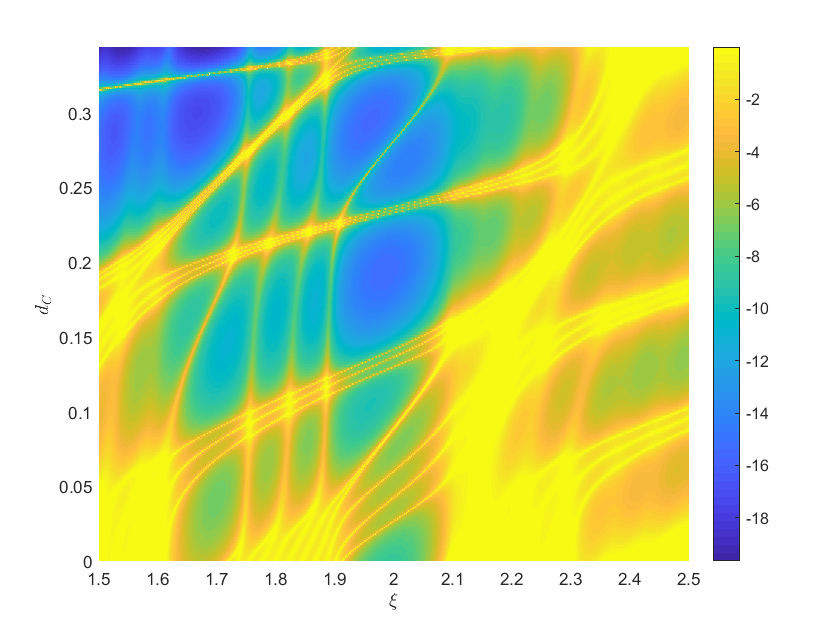}	      \label{Trans_bqb_M}}
	\caption{Transmission map for a double interface heterostructure. Layer $C$ of the quaternary PC is metallic. Configuration is (a) $qbq$  (b)$bqb$}
	\label{Metal_Split_States}
\end{figure}

\begin{figure}[!ht]
	\centering
	\subfloat[][]{\includegraphics[width=0.4\columnwidth]{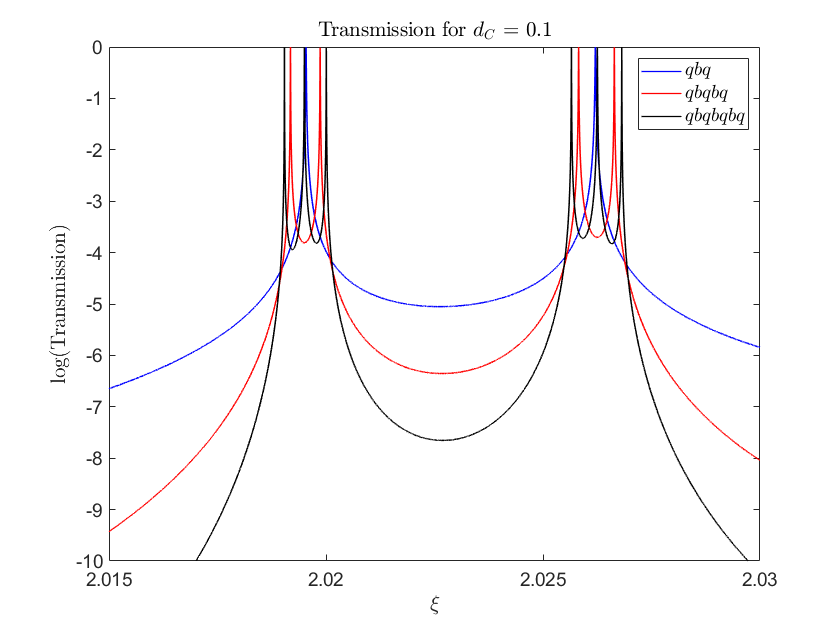}   \label{Split_metal} }
	\subfloat[][]{\includegraphics[width=0.4\columnwidth ]{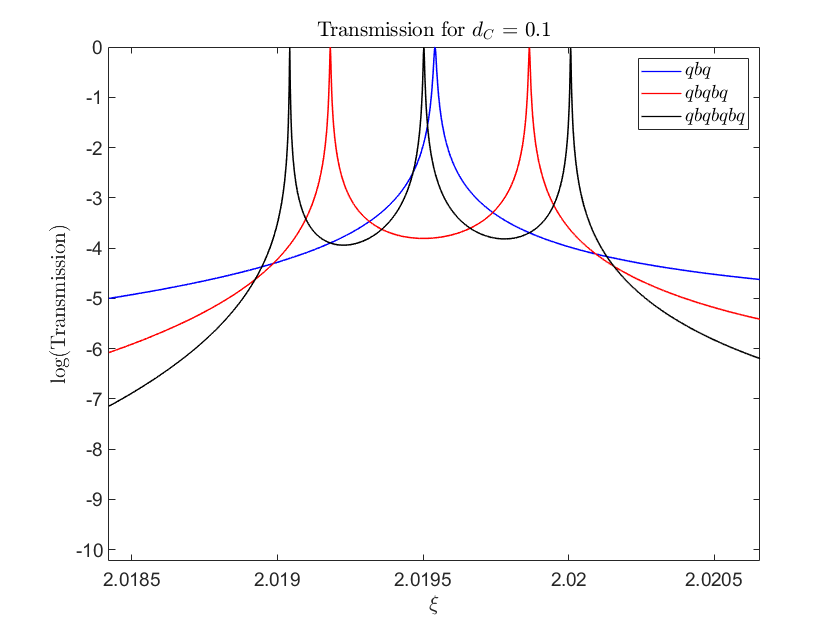}	      \label{Split_metal_zoom}}\\
	\subfloat[][]{\includegraphics[width=0.4\columnwidth]{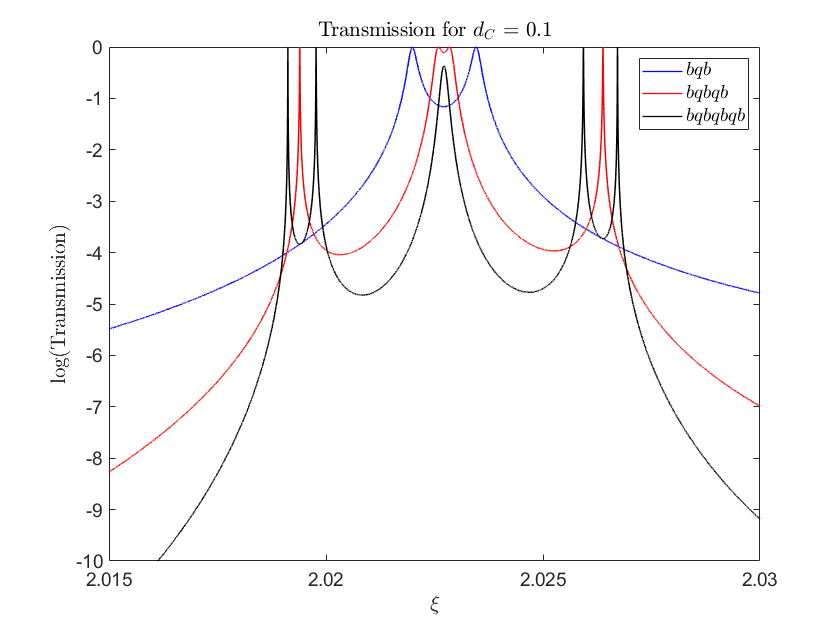}   \label{Split_metal2} }
	\caption{Transmission cross section for the heterostucture in Fig.~\ref{PCM} with layer $C$ having frequency dependence described by Eq.~\ref{Drude}. All other $\epsilon$ and $\mu$ values are the same as in Fig.~\ref{Trans}. (a) Transmission for structures of form $qb...bq$. (b) Zoomed in version of left collection of peaks in (a). (c) Transmission for structures of form $bq...qb$.}
	\label{Cross_Metal}
\end{figure}

As in the all dielectric case, when there are multiple binary/quaternary interfaces, topological states can split; however, the split states are much closer together, meaning that they are more difficult to resolve. Transmission maps for the $qbq$ and $bqb$ configurations are displayed in Fig.~\ref{Metal_Split_States}. While they look very similar to each other and to the single interface system, some subtleties can be pointed out. Resonances in the $qbq$ system are much sharper compared to those in $bqb$. Also the splitting can be seen, although it is more pronounced in $qbq$. Cross sections of the lower center topological state for $d_C=0.1$ are shown in both structures in Fig.~\ref{Cross_Metal} as the number of interface increases. In  Fig.~\protect\subref*{Split_metal}, the transmission is shown for a heterostructures sandwiched between two quaternary PCs. As the number of interfaces increases, each split state itself divides such that the total number equals the number of interfaces. Fig.~\protect\subref*{Split_metal_zoom} zooms into the left cluster of states. If the heterostructure is bounded by binary PCs, shown in Fig.~\protect\subref*{Split_metal2}, the two central split states appear much closer together. As the number of interfaces increases these two eventually merge and the resultant peak decreases. In the plot, this occurs for six interfaces ($bqbqbqb$).This makes it appear that there is a missing state; however, similar to the all dielectric heterostructures $qb...bq$ (Figs.~\protect\subref*{T_qbq}-\protect\subref*{T_qbqbqbq}), this merging occurs at lower values of $d_C$ as the number of interfaces increases. Therefore if the transmission cross section was taken for, say, $d_C=0.08$ rather than for $d_C=0.1$, then the central peak for structure $bqbqbqb$ would instead appear as a small doublet, bringing the total number of states to six.

\begin{figure}[!ht]
	\centering
	\subfloat{\includegraphics[width=0.5\columnwidth]{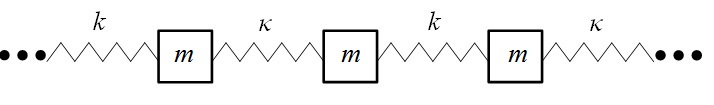}}
	\caption{Analogous coupled oscillator model of the heterostructure shown in Fig~\protect\ref{PCM}. The spring constants represent the PCs and the masses represent interfaces between PCs.}
	\label{OM}
\end{figure}

\begin{figure}[!ht]
	\centering
	\subfloat{\includegraphics[width=0.5\columnwidth]{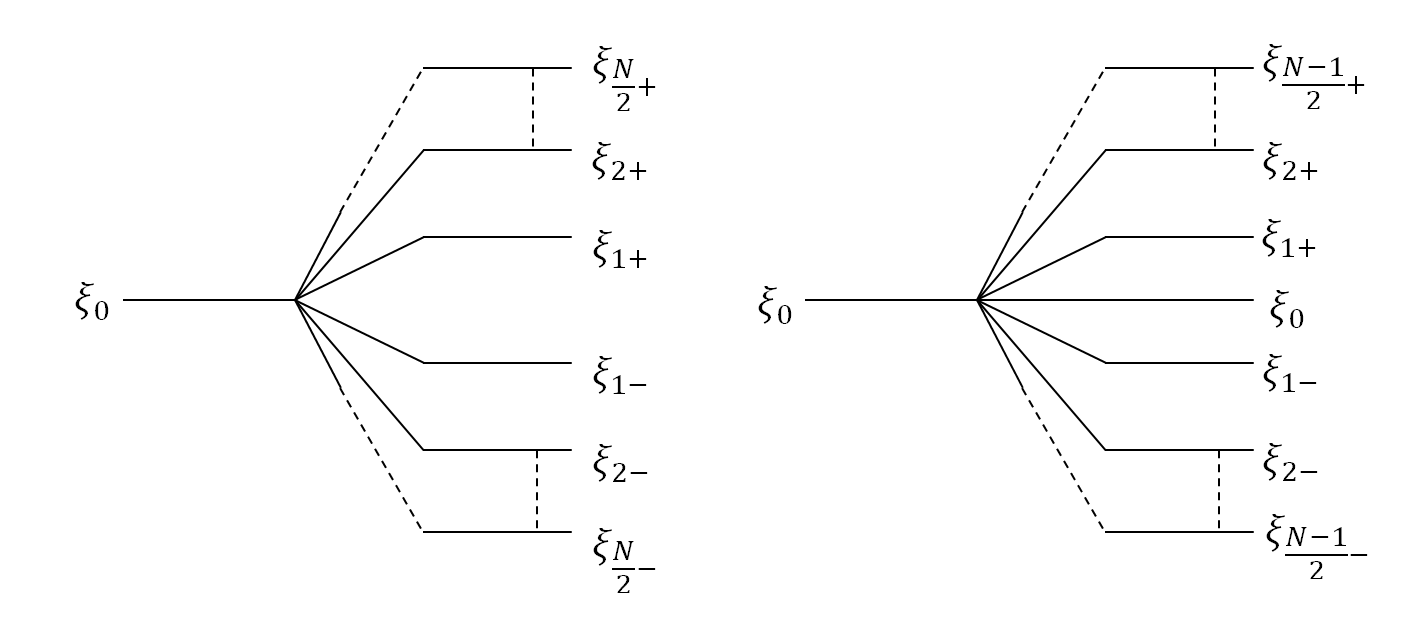}}
	\caption{Schematic showing the splitting of the original interface state. The number of interfaces is $N$. The left diagram represents an even number of interfaces while the right one is for an odd number of interfaces. }
	\label{Split}
\end{figure}

To help understand what is happening within the heterostructure, it is benefical to compare the optical system to the more familar 1D coupled harmonic oscillator, shown in Fig.~\ref{OM}. The interfaces between the individual PCs act as identical masses and the PCs themselves can be thought of as the spring constants. Since there are two different PCs, two distinct spring constants are used. In this example, the constant $k$ corresponds to the binary PC while $\kappa$ corresponds to the quaternary PC or vice virsa. The topological state will split into a number of states corresponding to the number of interfaces. With an even number of interfaces, the central state vanishes and splits such that half are above the original frequency and half are below. Using this analogy with two interfaces, the lower of the two states is the symmetric state while the higher one is the antisymmetric state \cite{Thornton}. With an odd number of interfaces, the central state still splits as in the even case except now the original state remains.This splitting is shown in Fig.~\ref{Split}. Overall, the original and split frequencies can be related by an average,

\begin{equation}
\xi_0^{2}=\frac{1}{N}\sum_{i=1}^{N} \xi_i^{2}  \label{xi_ave}
\end{equation}
where $N$ is the number of PCs in the entire structure and the index, $i$, is summed through all frequencies after the splitting.

\section{Conclusion}

We have described the evolution of resonant states in a photonic heterostructure composed of alternating binary and quaternary inversion symmetric photonic crystals as the quaternary crystal transforms from one binary to another. This was done by making the tunable layer in the quaternary crystal a free parameter. Two different heterostructures were given, one in which all components were dielectrics and one in which the free parameter was frequency dependent. For the all dielectric case, as shown in Fig.~\ref{Trans}, the maximum number of resonant states in a PBG is equal to the number of PC interfaces for all configurations shown. All configurations (except Figs.~\protect\subref*{T_qbq} and ~\protect\subref*{T_bq}) possess resonances that started as pass band states for $d_C=0$, but the central topological state can vanish, split in two, or remain intact, depending on if the heterostructure is $bq...qb$, $qb...bq$, or $bq...bq$ respectively. In Fig.~\ref{Metal_Split_States}, it was shown that if the additional layer in the quaternary PC has frequency dependence, some of the split modes can become much more compressed and difficult to resolve, more so for $bqb$ than $qbq$. Without a constant optical path, the topological states in Figs.~\ref{Trans_Impedance_bq_M} and \ref{Metal_Split_States} do not start and end at PBG closing points, but rather on the edges. These results show that it is possible to generate sequences of resonant states in a binary-quaternary PC heterostructure solely through the manipulation of the geometry of a heterostructure. The fabrication of such a structure could be useful for filtering applications.


\newpage

\begin{thebibliography}{100}

\bibitem{Yablonovitch}
E. Yablonovitch,
\emph{Inhibited Spontaneous Emission in Solid-State Physics and Electronics},
Phys. Rev. Lett. \textbf{58}, 2059 (1987).

\bibitem{John}
S. John,
\emph{Strong localization of photons in certain disordered dielectric superlattices},
Phys. Rev. Lett. \textbf{58}, 2486 (1987).

\bibitem{Istrate1}
E. Istrate, M. Charbonneau-Lefort, and E. H. Sargent,
\emph{Theory of photonic crystal heterostructure},
Phys. Rev. B \textbf{66}, 075121 (2002).

\bibitem{Istrate2}
E. Istrate, and E. H. Sargent,
\emph{Photonic crystal heterostructures and interfaces},
Rev. Mod. Phys. \textbf{78}, 455 (2006).

\bibitem{Tamm}
I. Tamm,
\emph{{\"U}ber eine m{\"o}gliche Art der Elektronenbindung an Kristalloberfl{\"a}chen},
Z. Phys. \textbf{76}, 849 (1932).

\bibitem{Vinogradov1}
A. P. Vinogradov, A. V. Dorofeenko, S. G. Erokhin, M. Inoue, A. A. Lisyansky, A. M. Merzlikin, and A. B. Granovsky,
\emph{Surface state peculiarities in one-dimensional photonic crystal interfaces},
 Phys. Rev. B \textbf{74}, 045128 (2006).

\bibitem{Vinogradov2}
A. P. Vinogradov, A. V. Dorofeenko, A. M. Merzlikin, and A. A. Lisyansky,
\emph{Surface states in photonic crystals},
Sov. Phys.-Usp \textbf{53}, 243 (2010).

\bibitem{Kavokin}
A. V. Kavokin, I. A. Shelykh, and G. Malpuech,
\emph{Lossless interface modes at the boundary between two periodic dielectric structures},
Phys. Rev. B \textbf{72}, 233102 (2005).

\bibitem{Gao}
D. Gao, W. Mao, R. Zhang, J. Liu, Q. Zhao, W. Y. Tam, and X. Wang,
\emph{Tunable interface state in one dimensional composite photonic structure}
Opt. Comm. \textbf{453}, 124324 (2019)

\bibitem{Lin}
L. L. Lin, and Z. Y.  Li,
\emph{Interface states in photonic crystal heterostructures},
Phys. Rev. B \textbf{63}, 033310 (2001).

\bibitem{Feng}
S. Feng, H. Y. Sang, Z. Y. Li, B. Y. Cheng, and D. Z. Zhang,
\emph{Sensitivity of surface states to the stack sequence of one-dimensional photonic crystals},
J. Opt. A: Pure Appl. Opt. \textbf{7}, 374 (2005)

\bibitem{Zheng}
Y. Zheng, Y. Wang, J. Lou, and P. Xu,
\emph{Optical Tamm states in photonic structures made of inhomogeneous material},
Opt. Comm. \textbf{406}, 103 (2018)

\bibitem{Abouti}
O. El Abouti, E. H. El Boudouti, Y. El Hassouani, A. Noual, and B. Djafari-Rouhani,
\emph{Optical Tamm states in one-dimensional superconducting photonic crystal},
Phys. Plasmas \textbf{23}, 082115 (2016)

\bibitem{Wang}
T. B. Wang, C. P. Yin, W. Y. Liang, J. W. Dong, and H. Z. Wang,
\emph{Electromagnetic surface modes in one-dimensional photonic crystals with dispersive metamaterials},
J. Opt. Soc. Am. B \textbf{26}, 1635 (2009)

\bibitem{Barvestani}
J. Barvestani, M. Kalafi, A. Soltani-Vala, and A. Namdar,
\emph{Backward surface electromagnetic waves in semi-infinite one-dimensional photonic crystals containing left-handed materials},
Phys. Rev. A \textbf{77}, 013805 (2008)

\bibitem{Namdar}
A. Namdar, I. V. Shadrivov, and Y. S. Kivshar,
\emph{Backward Tamm states in left-handed metamaterials},
Appl. Phys. Lett. \textbf{89}, 114104 (2006)

\bibitem{Hajian}
H. Hajian, B. Rezaei, A. S. Vala, and M. Kalafi,
\emph{Tuned switching of surface waves by a liquid crystal cap layer in one-dimensional photonic crystals},
Appl. Optics \textbf{51}, 2909 (2012)

\bibitem{Bashiri}
J. Bashiri, B. Rezael, J. Barvestani, and C. J. Zapata-Rodriguez,
\emph{Bloch surface waves engineering in one-dimensional photonic crystals with a chiral cap layer},
J. Opt. Soc. Am. B \textbf{36}, 2106 (2019)

\bibitem{Kaliteevski}
M. Kaliteevski, I. Iorsh, S. Brand, R. A. Abram, J. M. Chamberlain, A. V. Kavokin, and I. A. Shelykh,
\emph{Tamm plasmon-polaritons: Possible electromagnetic states at the interface of a metal and a dielectric Bragg mirror},
Phys. Rev. B \textbf{76}, 165415 (2007)

\bibitem{Brand}
S. Brand, M. A. Kaliteevski, and R. A. Abram,
\emph{Optical Tamm states above the bulk plasma frequency at a Bragg stack/metal interface},
Phys. Rev. B \textbf{79}, 085416 (2009)

\bibitem{Zhou}
H. Zhou, G. Yang, K. Wang, H. Long, and P. Lu,
\emph{Multiple optical Tamm states at a metal-dielectric mirror interface},
Opt. Lett. \textbf{35}, 4112 (2010)

\bibitem{Xiao}
M. Xiao, Z. Q. Zhang, and C. T. Chan, 
\emph{Surface Impedance and Bulk Band Geometric Phases in One-Dimensional Systems},
Phys. Rev. X \textbf{4}, 021017 (2014)

\bibitem{Zak}
J. Zak,
\emph{Berry's Phase for Energy Bands in Solids},
Phys. Rev. Lett. \textbf{62}, 2747 (1989)

\bibitem{Choi}
K. H. Choi, C. W. Ling, K. F. Lee, Y. H. Tsang, and K. H. Fung,
\emph{Simultaneous multi-frequency topological edge modes between one-dimensional photonic crystals},
Opt. Lett. \textbf{41}, 1644 (2016)

\bibitem{Cai}
L. Wang, W. Cai, M. Bie, X. Zhang, and J. Xu,
\emph{Zak phase and tolopological plasmonic Tamm states in one-dimensional plasmonic crystals},
Opt. Express \textbf{26}, 28963 (2018)

\bibitem{Yang}
Y. Yang, T. Xu, Y. F. Xu, and Z. H. Hang,
\emph{Zak phase induced multiband waveguide by two dimensional photonic crystals},
Opt. Lett. \textbf{42}, 3085 (2017)

\bibitem{Fei}
Y. Fei, Y. Liu, D. Dong, K. Gao, S. Ren, and Y. Fan,
\emph{Multiple adjustable optical Tamm states in one-dimensional photonic quasicrystals with predisigned bandgaps},
Opt. Express \textbf{26}, 34872 (2018)

\bibitem{Iorsh}
I. Iorsh, P. V. Panicheva, I. A. Slovinskii, and M. A. Kaliteevski,
\emph{Coupled Tamm Plasmons},
Tech. Phys. Lett. \textbf{38}, 351 (2012) 

\bibitem{Durach}
M. Durach, and A. Rusina,
\emph{Transforming Fabry-Perot resonaces into a Tamm mode},
Phys. Rev. B \textbf{86}, 235312 (2012)

\bibitem{Cox}
J. D. Cox, J. Sabarinathan, and M. R. Singh,
\emph{Resonant Photonic States in Coupled Heterostructure Photonic Crystal Waveguides},
Nanoscale Res. Lett. \textbf{5}, 741 (2010)

\bibitem{Hu}
J. Hu, W. Liu, W. Xie, W. Zhang, E. Yao, Y. Zhang, and Q. Zhan,
\emph{Strong coupling of optical interface modes in a 1D topological photonic crystal heterostructure/Ag hybrid system},
Opt. Lett. \text{44}, 5642 (2019)

\bibitem{Bianchi}
N. Bianchi, and L. Kahn,
\emph{Topological Photonic States at a 1-D Binary-Quaternary Interface},
 arXiv:1910.02920v1. 
(submitted for publication)

\bibitem{Yariv}
A. Yariv, and P. Yeh,
\emph{Optical Waves in Crystals: Propagation and Control of Laser Radiation},
(John Wiley \& Sons, 1984). 

\bibitem{Orfanidis}
S. J. Orfanidis,
\emph{Electromagnetic Waves and Antennas},
(Rutgers University, 2016).

\bibitem{Manzanares-Martinez}
J. Manzanares-Martinez,
\emph{ANALYTIC EXPRESSION FOR THE EFFECTIVE PLASMA FREQUENCY IN ONE-DIMENSIONAL METALLIC-DIELECTRIC PHOTONIC CRYSTAL},
Progress In Electromagnetic Research M \textbf{13}, 189 (2010).

\bibitem{Thornton}
S. T. Thornton, and J. B. Marion,
\emph{Classical Dynamics of Particles and Systems},
(Cengage Learning, 2008).

\end{thebibliography}
\end{document}